\RequirePackage{amsmath}
\documentclass[12pt]{iopart}
\usepackage{subfig}
\usepackage{graphicx}
\usepackage{amsmath}
\usepackage{amsfonts}
\usepackage{mathtools}
\usepackage{amssymb}
\usepackage{xcolor}

\usepackage[utf8]{inputenc}

\usepackage{newtxtext}
\usepackage{newtxmath}

\usepackage{cite}

\DeclareMathAlphabet{\mathcal}{OMS}{cmsy}{m}{n}

\def \Wc{\omega_{\textrm{con}}}
\def \Wd{\omega_{\textrm{gen}}}
\def \Wo{\omega_{\textrm{out}}}
\def \hh{\text{\scriptsize $\cal{I}$}}
\def \Ad{\mathcal{A}_{\textrm{gen}}}
\def \Ao{\mathcal{A}_{\textrm{out}}}

\def \Sc{\sigma_{\textrm{gen}}}
\def \Si{\sigma_{\textrm{int}}}

\begin{document}
\title{Thermodynamic cost of external control}
\author{Andre C. Barato$^1$ and Udo Seifert$^2$}
\address{$^1$ Max Planck Institute for the Physics of Complex Systems, N\"othnitzer Str. 38, 01187 Dresden, Germany\\
$^2$II. Institut f\"ur Theoretische Physik, Universit\"at Stuttgart, 70550 Stuttgart, Germany
}
\date{\today}

\begin{abstract}
Artificial molecular machines are often driven by the periodic variation
of an external parameter. This external control exerts work on the
system of which a part can be extracted as output if the system runs
against an applied load. Usually, the thermodynamic cost of the process
that generates the external control is ignored. Here, we derive a refined
second law for such small machines that include this cost, which is,
for example, generated by free energy consumption of a chemical reaction
that modifies the energy landscape for such a machine. In the limit of
irreversible  control, this refined second law becomes the
standard one. Beyond this ideal limiting case, our analysis shows that
due to a new entropic term unexpected regimes can occur: The control
work can be smaller than the extracted work and the work required to
generate the control can be smaller than this control work. Our general
inequalities are illustrated by a paradigmatic three-state system.
\end{abstract}


\section{Introduction}

Thermodynamic systems driven by external periodic control that reach a periodic steady state constitute a main 
class of systems out of equilibrium. Such systems are also known as ``stochastic pumps'' \cite{astu11} or 
"pulsating ratchets" \cite{reim02a} in the context of Brownian motors. For these systems an external protocol 
for the periodic variation of energies and energy barriers can lead to a net current.
Recent theoretical results for such systems include no-pumping theorems \cite{raha08,cher08,maes10,mand14,asba15}, a general theoretical 
framework for systems with periodic temperature (and other parameters) variations \cite{bran15,proe15}, a mapping between periodic steady states and nonequilibrium steady 
states \cite{raz16}, the relation between cost and precision in Brownian clocks \cite{bara16a}, the analysis of stochastic protocols \cite{bara16a,verl14}, 
limits on thermodynamic efficiency \cite{raha11}, generation of current with a hidden pump \cite{espo15}, and the study of large fluctuations \cite{rost17}.

On the experimental side, synthetically made molecular machines constitute a promising field for 
future applications \cite{erba15}. In particular, net motion in a given direction due to external control
has been achieved in several experiments \cite{leig03,eelk06,li15}. Interestingly, more recently an autonomous synthetic molecular 
machine that leads to rotation of a small ring on a larger ring of a catenane 
has been realized experimentally \cite{wils16}. In this case, the control, i.e., the periodic change of energies and energy barriers,  
is exerted by bulky groups that can attach to and detach from the larger ring blocking transitions 
between a link. These chemical reactions leading to attachment and detachment consume free energy. Such an autonomous 
synthetic machine is more similar to biological motors, which, typically, consume ATP.

\begin{figure}
\centering\includegraphics[width=120mm]{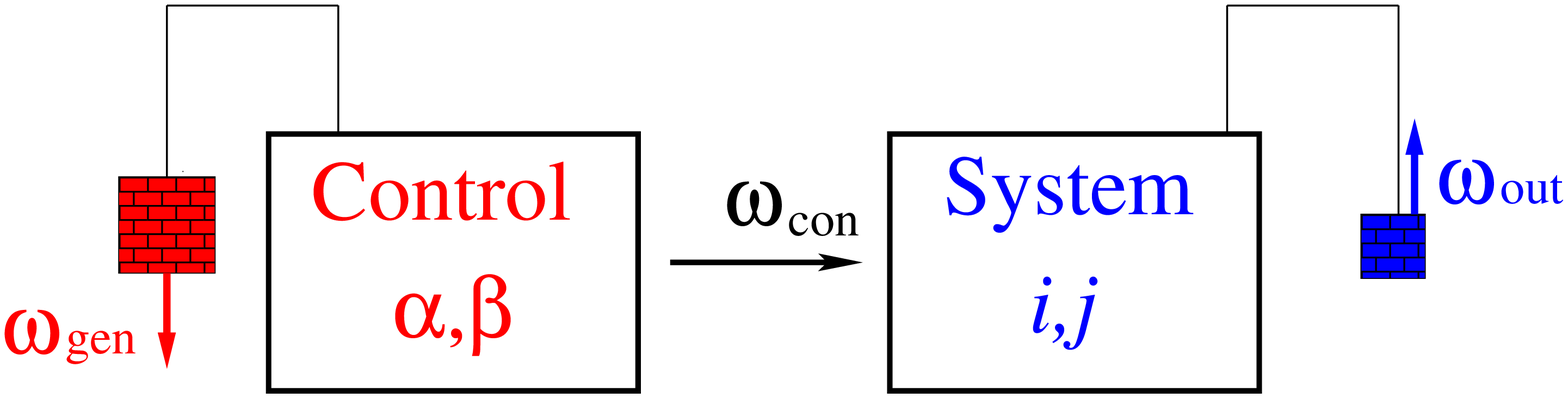}
\vspace{-3mm}
\caption{Representation of thermodynamically consistent external control. States of the external control are represented by $\alpha$ and $\beta$,
and states of the internal system are represented by $i$ and $j$. The power (work per time) to generate the external control $\Wd$ comes from a 
thermodynamic force, represented by the red falling square, that acts on the control states. The extracted power $\Wo$ is used to lift the blue 
square. The control power coming from the action of the controller on the system is  denoted $\Wc$.}
\label{fig1}
\end{figure}

In standard thermodynamics, the deterministic variation in time of an
external parameter leads to work exerted on the system. This control work is
given by the average change in the energy of the system due to changes of
the external parameter \cite{jarz96}. Part of this control work can be
extracted as output if an external load is applied to the system. For this
well known situation the energetic cost of generating the external control
does not appear in the second law. However, for an autonomous machine, 
illustrated in Fig. \ref{fig1}, where a thermodynamically consistent external
control is generated by the free energy difference in a chemical reaction, the 
second law has to include this cost. 

In this paper, we obtain generalized second law inequalities that incorporate
the cost of external control in a thermodynamically consistent way. Our
inequalities relate the work to generate the
external control, the work done on the system through external control, the extracted work,  
and an entropic term that quantifies correlations between the dynamics of the internal system and the state of the external control.

There is a particular limit, in which our new results have to become the known
inequality for systems driven by periodic control, which is the statement that 
the control work is larger than the extracted work. In this limit, which we call
the limit of irreversible control, the external control moves the parameters 
unidirectionally leading to a cost of external control that formally diverges. 
It is then reasonable to expect that the cost to generate the control 
is larger than the control work done on the system. Furthermore, due to the
statement of the second law for this known case of irreversible control, it is also
reasonable to expect that the extracted work is smaller than the control work.  We 
show that due to the presence of a new entropic term these expectations are not
necessarily correct for the realistic case of a thermodynamically consistent control,
which cannot be fully irreversible. The cost to 
generate the control can then be smaller than the control work and the extracted work
can be larger than the control work, which can be even negative. 

From a conceptual 
perspective, our results show that the standard periodically driven steady states can be seen as
a particular limit of a bipartite system. Indeed, our refined second law inequalities 
that account for the cost of external control follow from the theoretical framework
for bipartite systems \cite{bara13b,bara13a,hart14,horo14,bara14a}. 
We note that in a recent related study a bound on ``dissipation'' that considers
the entropic term to drive the external control has been obtained in \cite{mach15}.

The paper is organized in the following way. In Sec. \ref{sec2} we introduce our main result with a simple model. Our main 
result for the general setup is derived in Sec. \ref{sec3}. We illustrate our refined inequalities with a three-state model in Sec. \ref{sec4}. 
We conclude in Sec. \ref{sec5}.

\section{Illustrative Example}
\label{sec2}

\begin{figure}
\centering\includegraphics[width=100mm]{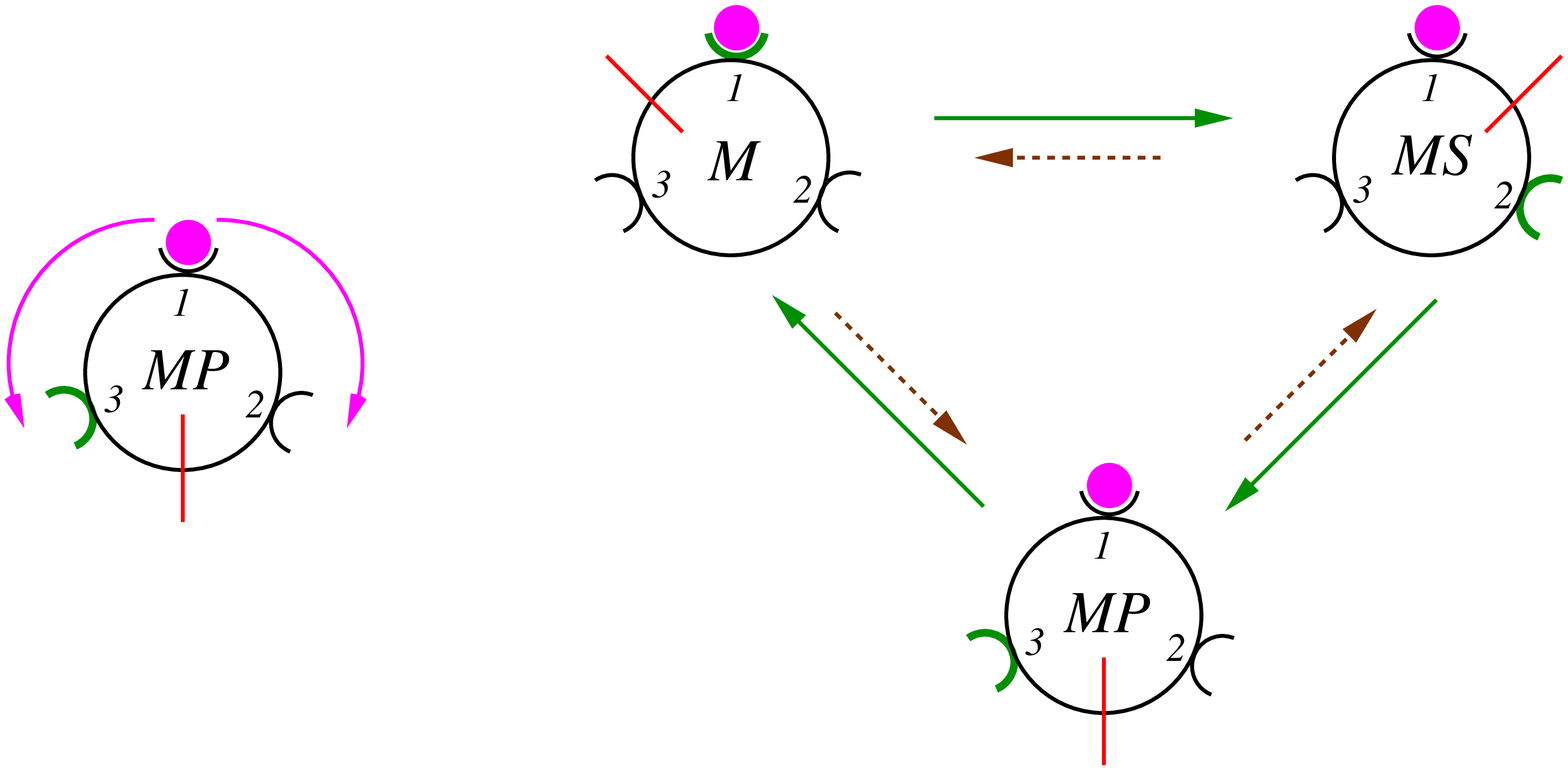}
\vspace{-3mm}
\caption{Three-state model. In the internal transition the magenta particle jumps between the three internal states $1,2,3$.
For example, if the particle sits at position $1$ and the red energy barrier is between $2$ and $3$, it can jump to either $2$ or $3$ 
in an internal transition, as shown on the left part of the figure. The green and brown arrows represent the external control 
that lead to changes in the energies and energy barriers of the internal system, as shown on the right. Without the dashed brown 
arrows, the control is irreversible. 
}
\label{fig2}
\end{figure}

Our main result, which is Eq. \eqref{seclaw2} below, can be illustrated with a simple model for a small machine driven by 
external control shown in Fig. \ref{fig2}. The magenta particle can be in three different positions, each representing a different state of the internal system. 
We assume that our model has Markovian dynamics with the particle jumping between these three positions. For instance, 
the three positions could be three different states of an enzyme $M_1$, $M_2$, and $M_3$, with the change between two states corresponding to 
a rotation of $120^{\circ}$ of the enzyme, similar to the case of F1-ATPase, see, e.g., \cite{zimm14} and references therein. The green position 
represents a state with energy $E$, whereas the other two black positions represent 
states with energy $0$. The red line represents an infinite energy barrier that does not allow transitions between the respective states. 
 
The external control is represented by the green arrows in Fig. \ref{fig2}. Changes in the control state leads to changes in the energies
of the internal states and in the energy barriers between internal states. These changes can happen at fixed times for a deterministic protocol or at 
exponentially distributed waiting times for a stochastic protocol \cite{verl14,bara16a}. An internal current, i.e., net movement of the particle in the circle, in the clockwise direction can be induced by the external 
control in Fig. \ref{fig2}. If the particle is moving against a load that leads to a thermodynamic force $\Ao$ in the anticlockwise direction, the 
system can do work against this force at a rate $\Wo$. This load would be the torque for an enzyme that rotates. The second law for this 
system with this irreversible control implies the inequality  
\begin{equation}
\Wc\ge \Wo,
\label{seclaw1}
\end{equation}
where $\Wc$ is the control power exerted on the internal system. 

For a stochastic protocol, the green arrows represent transitions between states
of the full system composed of the three-state ring and the external control. We call irreversible 
control the limit for which the transition rates represented by the brown dotted arrows in 
Fig. \ref{fig2} vanish. However, if we want to have a protocol that is thermodynamically  
consistent, we have to consider the possibility of reversed transitions. A physical model 
for this external protocol is  an enzyme $M$ driven by the chemical potential difference between 
a substrate $S$ and a product $P$, where $S$ could be ATP. For instance, the external 
transitions related to the green arrows in Fig. \ref{fig2} can lead to the cycle $M_i+S\to M_iS\to M_iP\to  M_i+P$, where $i=1,2,3$.
This cycle is then driven by the affinity $\Ad= (\varDelta\mu)/(k_BT)\ge 0$, where $\varDelta\mu=\mu_S-\mu_P$ is the chemical potential difference,
$k_B$ is Boltzmann's constant and $T$ is the temperature. When the molecule binds the substrate ($M_i+S\to M_iS$), or transforms  the substrate into product ($M_iS\to M_iP$), or releases 
a product in the solution ($M_iP\to M_i+P$), the energies and energies barriers between the internal states $i$ change. These energies also 
change with the respective reversed transitions. 

The rate of chemical work $\Wd$, which comes from free energy consumption due to the net transformation of $S$ into $P$, is the cost (per time) 
to generate the external control. Part of this power is transformed into mechanical power $\Wo$ against the 
torque $\Ao$ in the anti-clockwise direction. The standard second law of thermodynamics for this full thermodynamically consistent 
system composed of the internal system and the external control states reads \cite{seif12}  
\begin{equation}
k_BT\sigma=\Wd-\Wo\ge 0,
\label{seclaw15}
\end{equation}
where $\sigma$ is the rate of entropy production (defined without $k_B$). The important point here is that if we compare this inequality 
with Eq. \eqref{seclaw1}, we see that the term $\Wc$ has disappeared. Hence, in the limit of irreversible control that leads 
to $\Wd\to \infty$, Eq. \eqref{seclaw15} does not become the well known Eq. \eqref{seclaw1}.

What are the inequalities that generalize Eq. \eqref{seclaw1} for this case of a thermodynamically consistent 
protocol? As derived below, the answer will be given by the following refined second law inequalities,
\begin{equation}
\Wd\ge \Wc-k_BT\hh\ge \Wo,
\label{seclaw2}
\end{equation}
where $\hh$ is an entropic rate that quantifies correlations between the dynamics of the internal system and the state of the external control. 
In the limit of irreversible control, Eq. \eqref{seclaw2} reduces to $\Wc-k_BT\hh\ge \Wo$, where the informational term 
fulfills $\hh\ge 0$ in this limit of irreversible control. Eq. \eqref{seclaw2} is thus  a generalization  
of Eq. \eqref{seclaw1} that is also valid for the case of thermodynamically consistent control. 
In the next section, we derive this new refined second law as a direct consequence of the second law inequalities for 
bipartite processes from \cite{hart14,horo14}.


\section{General Theory}
\label{sec3}

\subsection{Second Law for the Full System}

We consider a bipartite Markov process in a stationary state \cite{bara13b,bara13a,hart14,horo14,bara14a}. States of the internal system are denoted by Roman letters $i$ and $j$ and 
states of the external protocol by Greek letters $\alpha$ and $\beta$. The transition rate from state $(i,\alpha)$ to state 
$(j,\beta)$ is
\begin{equation}
w_{ij}^{\alpha\beta}\equiv\left\{
\begin{array}{ll} 
 w^{\alpha\beta}_i & \quad \textrm{if $i=j$ and $\alpha\neq\beta$}, \\
 w^{\alpha}_{ij} & \quad  \textrm{if $i\neq j$ and $\alpha=\beta$},\\
 0 & \quad \textrm{if $i\neq j$ and $\alpha\neq\beta$}. 
\end{array}\right.\,
\label{defrates}
\end{equation}
Transitions that change the state of the internal system and the state of the external 
protocol simultaneously are not allowed.

The transition rates are related to the free energies and thermodynamic affinities. The free energy of an state $(i,\alpha)$ is denoted $F_i^\alpha$,
the affinity that drives the external protocol is denoted $\Ad$ and the affinity of the internal process is denoted $\Ao$.
The generalized detailed balance relation \cite{seif12} for transitions that change the internal states reads
\begin{equation}
\ln\frac{w^\alpha_{ij}}{w^\alpha_{ji}}= F_i^\alpha -F_j^\alpha- \Ao d_{ij},
\end{equation}
where we are assuming an isothermal system with $k_BT=1$ throughout. For transitions that change the external protocol, this relation is
\begin{equation}
\ln\frac{w^{\alpha\beta}_{i}}{w^{\beta\alpha}_{i}}= F_i^\alpha -F_i^\beta+ \Ad d^{\alpha\beta}.
\label{gendetext}
\end{equation}
The quantities $d_{ij}$ and $d^{\alpha\beta}$ are generalized distances. If $\Ad$ is a chemical 
potential difference, then $d^{\alpha\beta}$ is the number of substrate molecules consumed in the transition 
from $\alpha$ to $\beta$. If $\Ao$ is a torque, then $d_{ij}$ is an angle difference between $i$ and $j$.

The power required to generate the control is defined as  
\begin{equation}
\Wd\equiv \Ad\sum_i\sum_{\alpha<\beta} J_i^{\alpha\beta}d^{\alpha\beta},
\label{defwdri}
\end{equation}
where $J_i^{\alpha\beta}\equiv P_i^{\alpha}w_i^{\alpha\beta}-P_i^{\beta}w_i^{\beta\alpha}$ and the sum $\sum_{\alpha<\beta}$
is over all external links. The rate of extracted work is given by
\begin{equation}
\Wo\equiv \Ao\sum_\alpha\sum_{i<j} J^{\alpha}_{ij}d_{ij},
\label{defwout}
\end{equation}
where $ J^{\alpha}_{ij}\equiv P_i^\alpha w^{\alpha}_{ij}-P_j^\alpha w^{\alpha}_{ji}$ and the sum $\sum_{i<j}$ is over all 
internal links. The entropy production of the full system is
\begin{equation}
\sigma\equiv \sum_{i,\alpha}P_i^\alpha \sum_{\beta\neq\alpha} w^{\alpha\beta}_i\ln \frac{w^{\alpha\beta}_i}{w^{\beta\alpha}_i}+\sum_{i,\alpha}P_i^\alpha\sum_{j\neq i} w^\alpha_{ij}\ln\frac{w^\alpha_{ij}}{w^\alpha_{ji}}
=\Wd-\Wo\ge0,
\label{totalent}
\end{equation}
where the second equality follows from Eqs. \eqref{defwdri} and \eqref{defwout}. The inequality above is the standard second law from stochastic thermodynamics 
for the full bipartite process. In this paper, we restrict to the case $\Wd\ge0$. If $\Wd$ is negative, then the internal system plays the role of an external protocol and the external protocol 
plays the role of an internal system.

A key quantity for a system driven by external control that does not appear in this standard second law \eqref{totalent}
is the rate of work done on the system by external control, i.e., the control power
\begin{equation}
\Wc\equiv \sum_i\sum_{\beta<\alpha} J_i^{\alpha\beta}(F_i^\beta-F_i^\alpha)=\sum_\alpha\sum_{j<i} J_{ij}^{\alpha}(F_i^\alpha-F_j^\alpha).
\label{defcontrolw}
\end{equation}
The second equality comes from the conservation law $\frac{d}{dt}\sum_{i\alpha}F_i^\alpha P_i^\alpha=0$. 
Hence, $\Wc$ does not appear in the entropy production \eqref{totalent} because the terms leading to 
free energy changes  due to jumps that change the external control cancels the terms due 
to internal jumps.

\subsection{Refined Second Law}

The external protocol and internal system are two subsystems that form the full system. For a bipartite system, 
there are also second law inequalities for these subsystems \cite{hart14,horo14}.
The rate of entropy production associated only with the jumps that change the external protocol is given by
\begin{equation}
\Sc\equiv \sum_{i,\alpha}P_i^\alpha \sum_{\beta\neq\alpha} w^{\alpha\beta}_i\ln \frac{w^{\alpha\beta}_iP_i^\alpha}{w^{\beta\alpha}_iP_i^\beta}\ge 0.
\end{equation}
 Using Eqs. \eqref{defwdri} and \eqref{defcontrolw}, this second law 
for the external control alone reads  
\begin{equation}
\Sc= \Wd-\Wc+\hh\ge0,
\label{seclawx}
\end{equation}
where 
\begin{equation}
\hh\equiv \sum_{i}\sum_{\beta<\alpha} J_i^{\alpha\beta}\ln \frac{P^{\alpha}_i}{P^{\beta}_i}.
\end{equation}
This entropic rate is the rate at which jumps of the external control decrease the static mutual information (or increase the conditional Shannon entropy) \cite{hart14,horo14,hart16}.
If $\hh$ is positive, then the dynamics of the external control decreases the correlation between the subsystems. Bipartite systems have the following entropic conservation
law: the rate at which jumps of the internal system increase the static mutual information is exactly $\hh$. If $\hh$ is negative, the dynamics of the internal system 
decreases the correlation between the subsystems.

The rate of entropy production due to jumps related to the internal system is
\begin{equation}
\Si\equiv \sum_{i,\alpha}P_i^\alpha\sum_{j\neq i} w^\alpha_{ij}\ln\frac{w^\alpha_{ij}P_i^\alpha}{w^\alpha_{ji}P_j^\alpha}\ge 0
\end{equation}
For the internal subsystem the second law reads
\begin{equation}
\Si=\Wc-\hh-\Wo\ge 0,
\label{seclawy}
\end{equation}
where we used Eqs. \eqref{defwout} and \eqref{defcontrolw}. With Eqs. \eqref{seclawx} and \eqref{seclawy}, we obtain 
\begin{equation}
\Wd\ge \Wc-\hh\ge\Wo,
\label{seclawfinal}
\end{equation}
which is our refined second law in Eq. \eqref{seclaw2}. We note that here we consider an internal affinity $\Ao$ that is independet 
of $n$. For the case of several internal affinities that can depend on the external control, which is the case of a model 
that displays a phenomena known as negative mobility \cite{eich02,eich02a}, there will be different terms from those terms 
contained in Eq. \eqref{seclawfinal}. In principle, our theoretical framework should be generalizable 
to a case where the cost of such external control would be relevant.

\subsection{Limit of irreversible control}

The external protocol becomes unaffected  by the dynamics of the internal system in the following limit \cite{verl14}. The free energy difference  
is written as 
\begin{equation}
F_i^\beta-F_i^\alpha= E_\beta-E_\alpha+E^*_{\beta,i}-E^*_{\alpha,i},
\label{difflimit}
\end{equation}
where $E^*_\alpha$ is the energy of the state $\alpha$ of the external protocol and $E^*_{\alpha,i}$ is the interaction energy. 
From the generalized detailed balance 
relation \eqref{gendetext} we obtain $w^{\alpha\beta}_i\simeq w^{\alpha\beta}$, if this energy difference 
fulfills $E_\beta-E_\alpha\gg E^*_{\beta,i}-E^*_{\alpha,i}$. For such transition rates the external protocol alone 
is a Markov process with dynamics unaffected by the state of the internal system.

Even though $w^{\alpha\beta}_i\simeq w^{\alpha\beta}$, we have to account for the contribution coming from the interaction energy difference
in the inequalities \eqref{seclawfinal}. In particular, using Eq. \eqref{difflimit}, the control power in Eq. \eqref{defcontrolw} becomes 
\begin{align}
\Wc= \sum_{\beta<\alpha} J^{\alpha\beta}(E_{\beta}-E_{\alpha})+\sum_i\sum_{\beta<\alpha} J_i^{\alpha\beta}(E^*_{\beta,i}-E^*_{\alpha,i})=\sum_i\sum_{\beta<\alpha} J_i^{\alpha\beta}(E^*_{\beta,i}-E^*_{\alpha,i}).
\end{align}
where $J^{\alpha\beta}= \sum_iJ_i^{\alpha\beta}$. The term $\sum_{\beta<\alpha} J^{\alpha\beta}(E_{\beta}-E_{\alpha})=0$ because the dynamics of the 
external protocol alone is Markovian.

The limit of irreversible  control corresponds to irreversible rates for jumps of the external protocol. If the external protocol has $N$ states, 
$\alpha=1,2,\ldots,N$ then we write the rates as $w^{\alpha,\alpha+1}=\gamma_\alpha$
and $w^{\alpha+1,\alpha}=0$, with $\alpha+1=1$ for $\alpha=N$. Such irreversible rates correspond to a formally divergent affinity $\Ad$ in 
\eqref{gendetext}, leading to $\Wd\to \infty$. Therefore, in this limit of irreversible control the refined second law \eqref{seclawfinal} leads to 
\begin{equation}
\Wc\ge\Wo+\hh\ge\Wo,
\label{seclawperfect}
\end{equation}
where we used the fact that $\hh\ge 0$ for transition rates $w^{\alpha\beta}$ independent of $i$ \cite{bara14a}.

\section{Three-state model and time-scale separation}
\label{sec4}

\subsection{Illustration of the refined inequalities}

We now consider a more general version of the model in Fig. \ref{fig2}, with arbitrary energies and energy barriers. The three internal states 
are three different rotation angles of the enzyme $i=1,2,3$ and the three states of the external protocol are $M_i$, $M_iS$, and $M_iP$, which correspond 
to $\alpha=1,2,3$, respectively. The total Markov process of internal system and external protocol together has then nine states.
The transition rates for an internal change are set to 
\begin{equation}
w^{\alpha}_{i,i+1}= k \textrm{e}^{F_i^\alpha-B_{i}^\alpha},
\end{equation}
for a clockwise rotation,
\begin{equation}
w^{\alpha}_{i+1,i}= k \textrm{e}^{\Ao/3}\textrm{e}^{F_{i+1}^\alpha-B_{i}^\alpha}
\end{equation}
for an anti-clockwise rotation. The quantities $B_{i}^\alpha$ represent energy barriries 
between states. The transition rates for a change in the external protocol are given by 
\begin{equation}
w^{\alpha,\alpha+1}_{i}= \gamma,
\end{equation}
and
\begin{equation}
w^{\alpha+1,\alpha}_{i}= \gamma \textrm{e}^{-\Ad/3}\textrm{e}^{F_i^{\alpha+1}-F_i^\alpha}.
\end{equation}
The parameter $k$ characterizes the speed of internal transitions and the parameter $\gamma$ characterizes the 
speed of changes in the external protocol.

\begin{figure}
\subfloat[][]{\includegraphics[width=75mm]{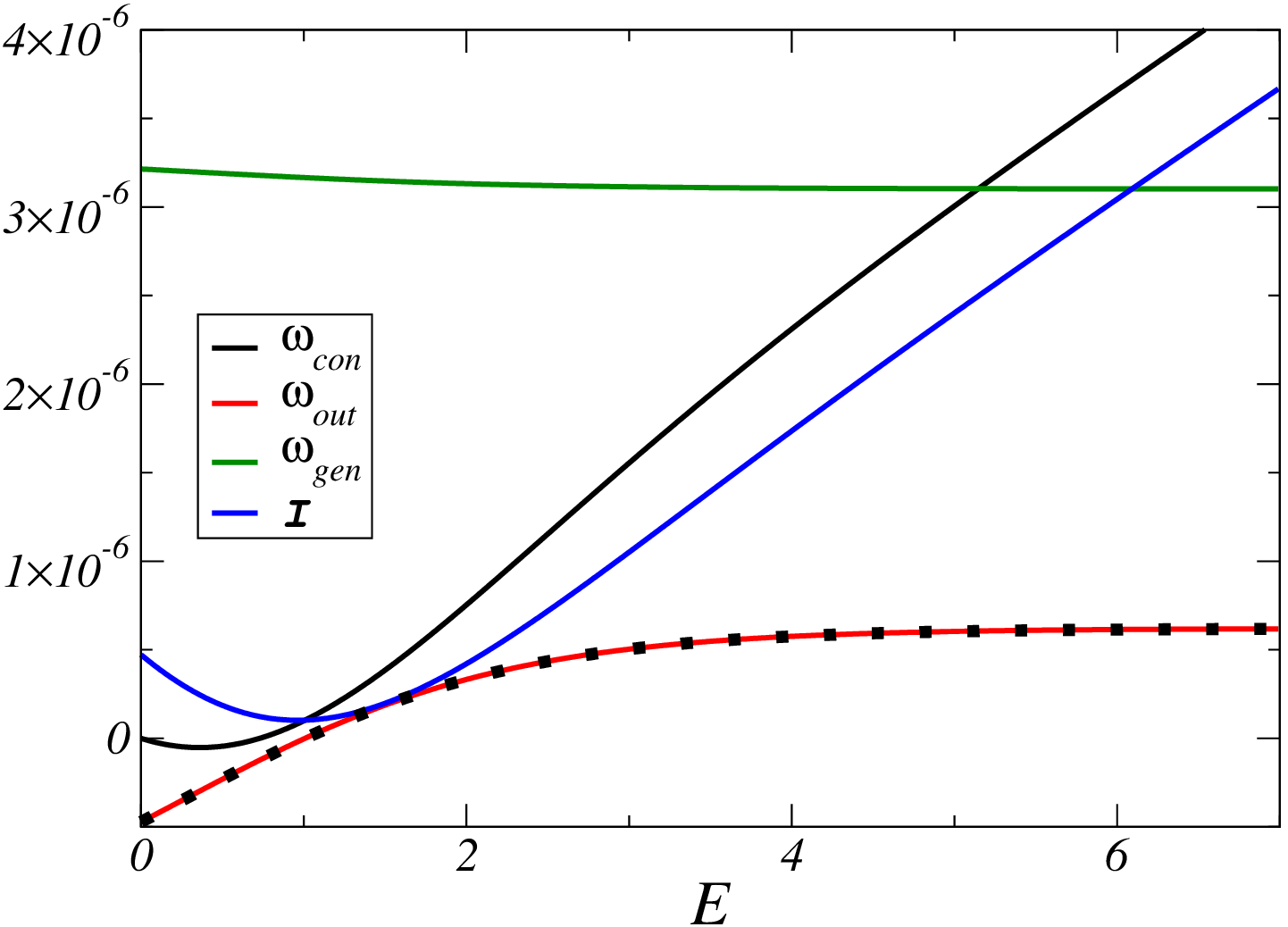}\label{fig3a}}
\subfloat[][]{\includegraphics[width=75mm]{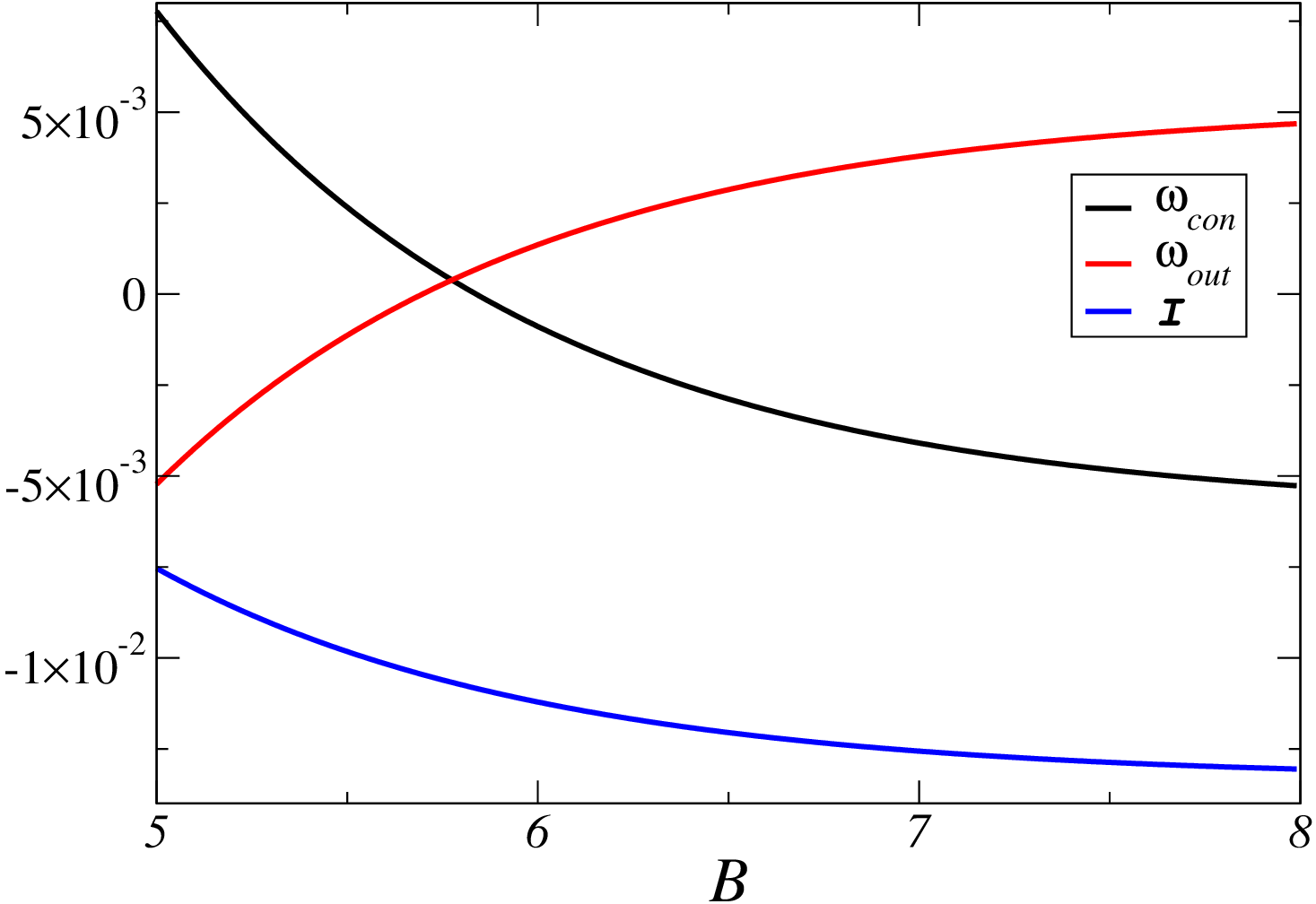}\label{fig3b}}
\vspace{-3mm}
\caption{Exact results for the three-state model.(a) The parameters are set to $\Ad= 10$, $\Ao= 2$, $F_1=E$, $B_3\to\infty$, $F_2=F_3=B_1=B_2=0$, $k=1$, $\gamma=10^{-6}$.
The black points, which match with the red curve $\Wo$, represent $\Wc-\hh$. (b) The parameters are set to $\Ad= 3.35$, $\Ao= 1.53$, $F_1=0.86$, $F_2=0.17$, $F_3=1.91$,
$B_1=-1.04$, $B_2=B$, $B_3=0.33$, $k=1$, and  $\gamma=\textrm{e}^{1.55}$.}
\label{fig3}
\end{figure}

A symmetric protocol is obtained with the energies and energy barriers given by
\begin{equation}
F_i^\alpha= F_{i-\alpha+1},
\end{equation}
and
\begin{equation}
B_i^\alpha= B_{i-\alpha+1}.
\end{equation}
where we assume periodic boundary conditions for the subscript $i-\alpha+1$. 
With this symmetric choice we can reduce the stochastic matrix for the full Markov process with dimension nine to a stochastic matrix 
with dimension three \cite{bara16a}. This reduction facilitates the analytical calculation that leads to a stationary distribution with
quite long expression in term of the parameters for the general case. The model shown in Fig. \ref{fig2} corresponds to the choice 
$F_1=E$, $B_3\to \infty$ and $F_2=F_3=B_1=B_2=0$.

In Fig. \ref{fig3} we illustrate the major role played by the entropic rate $\hh$ for a thermodynamically consistent external control. 
Due to this entropic contribution two somewhat surprising situations can happen. First,  in Fig. \ref{fig3a}, we show 
that the power to generate the external control $\Wd$ can be smaller than the control power exerted on the system $\Wc$. Second, 
in Fig. \ref{fig3b}, we show that the extracted power $\Wo$ can be larger than the control power $\Wc$, which can be negative.

From the second law inequalities in Eq. \eqref{seclawfinal} we can define the efficiencies $\eta\equiv \Wo/\Wd$, $\eta_{\textrm{con}}\equiv (\Wc-\hh)/\Wd$, and 
$\eta_{\textrm{int}}\equiv \Wo/(\Wc-\hh)$. The first efficiency $\eta$ is the standard efficiency for a nonequilibrium steady state \cite{seif} corresponding to the full
bipartite process. This efficiency compares the extracted power with the full cost to generate the external control.
The second efficiency $\eta_{\textrm{gen}}$ gives the fraction of the power to generate the external control that is transformed into 
control power minus the entropic rate $\hh$. Interestingly, in the limit of irreversible control  the ratio $\Wo/\Wc$ is 
an efficiency that quantifies the amount of the control work that is transformed into extracted work \cite{raha11}.
For the general case, $\Wo/\Wc$ becomes a pseudo-efficiency since it can be larger than one. For a thermodynamic consistent control
the third efficiency $\eta_{\textrm{int}}$ should rather be used to characterize the performance of the machine to convert ``control power''
into output power.

\subsection{Time-scale separation}

If there is time-scale separation then, with a few assumptions, we can show that the second law inequality \eqref{seclawy} for the internal subsystem
is saturated. The internal rates $w_{ij}^\alpha$ are assumed to be of order $k$ and the external rates $w_{i}^{\alpha\beta}$ 
are assumed to be of order $\gamma$, with $k\gg \gamma$. In this case, the power to drive the control $\Wd\ge0$ is of order $\gamma$. If we impose that 
$\Wo\ge0$, then from the standard second law \eqref{totalent}, $\Wo$ must also be of order $\gamma$. 

Since $\Wo$ is of order $\gamma$, it is reasonable to expect that the internal currents $J_{ij}^\alpha$ that appear in Eq. \eqref{defwout} are 
also of order $\gamma$. In this case, form Eq. \eqref{seclawy} we obtain
\begin{align}
\Wc-\hh-\Wo= \sum_{\alpha}\sum_{j<i} J_{ij}^{\alpha}\ln\frac{W_{ij}^\alpha P_i^\alpha}{W_{ji}^\alpha P_j^\alpha}=\sum_{\alpha}\sum_{j<i} J_{ij}^{\alpha}\ln\left(1+\frac{J_{ij}^{\alpha}}{W_{ji}^\alpha P_j^\alpha}\right)
=\gamma\textrm{O}\left(\frac{\gamma}{k}\right).
\label{eqsep}
\end{align}
Hence, in the limit where changes in the external protocol are infinitely slower than the internal transitions the second inequality in 
\eqref{seclawfinal} is saturated, i.e., $\Wc-\hh=\Wo$. This equality is illustrated with the three-state model in Fig. \ref{fig3a}.

The typical case of irreversible control with a deterministic protocol can be recovered if we consider a stochastic protocol with a large number of 
jumps $N$ and a rate $\gamma$ for a change of the external protocol that scales with $N$ \cite{bara16a,brit17}. In this case, the entropic 
rate $\hh$ goes to zero and we obtain $\Wc=\Wo$ with the separation of time scales in Eq. \eqref{eqsep}, a known result in thermodynamics.

\section{Conclusion}
\label{sec5}

We have obtained refined second law inequalities for machines driven by periodic external control that take the thermodynamic cost 
to generate the external control into account. Our inequalities establish a relation between the cost to generate external control, 
the control work exerted on the internal system, and the extracted work. In particular, we have shown that the cost for external 
control can be smaller than the control work and that the extracted work can be larger than the control work. 
These regimes result from the entropic term $\hh$ in Eq. \eqref{seclawfinal} that quantifies correlations between the dynamics of the internal 
system and the state of the external control, which has to be stochastic for a thermodynamic consistent control.

From a conceptual perspective we have shown that systems driven by external control that reach a periodic steady state, which form a major 
class of nonequilibrium systems, can be seen as a particular limit of a steady state of a bipartite process. In this limit of irreversible control, 
the cost of control diverges and we are left only with the second inequality in Eq. \eqref{seclawfinal}. This result 
further demonstrates the power of the theoretical framework for bipartite systems developed in \cite{hart14,horo14,bara14a}.

Our refined inequalities correspond to the appropriate statement of the second law for a machine driven by 
a thermodynamically consistent control. This kind of control occurs in particular if the system is driven 
by free energy consumption of a chemical reaction, as is the case of the catenane analyzed experimentally in 
\cite{wils16}. We expect our formalism to play an important role for understanding and optimizing the 
operation of such autonomously driven machines.






 \end{document}